\pdfoutput=1
\documentclass[a4paper,10pt,twocolumn,amsmath,amssymb,nofootinbib,%
superscriptaddress,groupedaddress,floats,floatfix,aps,prd]{revtex4-1}
\usepackage[british]{babel}
\usepackage[latin1]{inputenc}
\usepackage[T1]{fontenc}
\usepackage[final]{showkeys} 
\usepackage{graphicx,color}
\usepackage{enumerate}
\usepackage{booktabs}

\usepackage[pdftex,pdfusetitle]{hyperref}
\hypersetup{colorlinks=true,linkcolor=black,citecolor=black,filecolor=black,urlcolor=black,
pdfauthor={Mariano Cadoni, Matteo Ciulu and Matteo Tuveri}}
\usepackage[]{cleveref}
\Crefname{equation}{Eq.}{Eqs.}
\Crefname{figure}{Fig.}{Figs.}
\crefformat{plural}{#2eqs.~(#1)#3}

\def\be#1\ee{\begin{align}#1\end{align}}

\newcommand{\bea}{\begin{eqnarray}}
\newcommand{\eea}{\end{eqnarray}}

\def\a{\alpha}

\def\lb{\label}

\newcommand{\beq}{\begin{equation}}
\newcommand{\eeq}{\end{equation}}
\newcommand{\bal}{\begin{align}}
\newcommand{\eal}{\end{align}}

\def\L{{\cal L}}

\begin{document}

\title{Symmetries, Holography and Quantum Phase Transition in Two-dimensional Dilaton AdS Gravity}

\date{\today}

\author{Mariano Cadoni}\email{mariano.cadoni@ca.infn.it}
\author{Matteo Ciulu}\email{matteo.ciulu@gmail.com}
\author{Matteo Tuveri}\email{matteo.tuveri@ca.infn.it}
\affiliation{Dipartimento di Fisica, Universit\`a di Cagliari
\& INFN, Sezione di Cagliari\\
Cittadella Universitaria, 09042 Monserrato, Italy}

\begin{abstract}
We present a revisitation of the Almheiri-Polchinski dilaton gravity 
model from a two-dimensional (2D) bulk perspective.  
We describe a peculiar feature of the model, namely the 
pattern of conformal symmetry breaking using bulk Killing vectors, 
a covariant definition of mass and the flow between different 
vacua of the theory. We show that the effect of 
the symmetry breaking is both the generation of an infrared
scale (a mass gap) 
and to make local the Goldstone modes associated with the asymptotic 
symmetries of the 2D spacetime. In this way a non vanishing central 
charge is generated in the dual conformal theory, which accounts for 
the microscopic entropy of the 2D black hole. The use of covariant 
mass allows to compare energetically the two different vacua of the 
theory and to show that at zero temperature the vacuum with a constant 
dilaton is energetically preferred. We also translate in the bulk language 
several features of the dual CFT discussed by Maldacena et {\it al}. The 
uplifting of the 2D model to $(d+2)-$dimensional theories 
exhibiting hyperscaling violation is briefly discussed.

\end{abstract}

\maketitle%

\section{Introduction}

Two-dimensional (2D) dilaton gravity models have a long history 
(see Ref.~\cite{Grumiller:2002nm} for a review). They have been first 
proposed for studying quantum gravity in a simplified context~\cite{Jackiw:1984je}. 
Later, they have been developed along several different directions as 
an arena to understand gravity in a simplified setting, as effective 
description of the radial modes (the S-wave sector) of black 
holes~\cite{Strominger:1994tn} and as toy models for black hole evaporation 
and related information puzzles~\cite{Callan:1992rs}. 2D dilaton 
gravity models have been also used to explore the Anti de Sitter/Conformal field theory 
(AdS/CFT) correspondence 
in two dimensions~\cite{Cadoni:1998sg,Cadoni:1999ja,Cadoni:2000gm,Caldarelli:2000xk} 
and to investigate the microscopic origin of the Bekenstein-Hawking 
black hole entropy~\cite{Cadoni:1998sg}. 

Recently, a 2D AdS dilaton gravity model has been proposed 
by Almheiri and Polchinski (AP)~\cite{Almheiri:2014cka}. The main motivation 
behind this proposal was to understand the infrared (IR) behaviour of 
higher-dimensional black holes, which flow in the IR to an AdS$_2\times S^d$ 
spacetime. Among many others, this is for instance the case of  
charged Reissner-Nordstr\"{o}m black holes in General Relativity. The 
IR behaviour of these black holes is problematic for several reasons. 
In  fact, the $T=0$ extremal black hole 
is a zero temperature vacuum state with non vanishing entropy and the 
backreaction is so strong that there are no finite energy excitations 
above the vacuum~\cite{Maldacena:1998uz,Almheiri:2016fws,Preskill:1991tb}.

The AP dilaton gravity model coupled to a matter field $f$ is described 
by the action~\cite{Almheiri:2014cka} (we set the dimensionless Newton 
constant to $G_{2}=1/8\pi$) 
\begin{equation} \label{2dlagrangian:eq}
S= \int d^{2} x \left[ \frac{1}{2}\sqrt{-g} \left( \eta  R - V(\eta)\right) + {\cal L}_{m} \right]+   \int dt {\cal L}_{b},
\end{equation}
where $\eta$ is a scalar field (the dilaton). The matter and boundary 
Lagrangian ${\cal L}_{m},\,  {\cal L}_{b}$ are given by 
\begin{equation}		
\label{lagrangians}
 {\cal L}_{m}= - \frac{1}{8} \sqrt{-g} \, 
			 (\nabla f)^2,\,
			{\cal L}_{b} =  \sqrt{|h|} \, \eta {\cal K},
\end{equation}
 where $h_{i j}$ is the induced metric on the boundary and  
${\cal K}$ is the trace of the second fundamental form. The potential 
for the dilaton is
\begin{equation}		
\label{potential}
V(\eta)=2\lambda^2( \alpha^2 -  \eta).
\end{equation}
Notice that the potential contains a dimensionless parameter $\alpha^2$ 
and a parameter $\lambda$ with dimensions $[L]^{-1}$. 

The AP model extends the well-known Jackiw-Teitelboim (JT) 
model~\cite{Jackiw:1984je}, characterised by a simple homogeneous potential, 
 by including the constant term $2\lambda^2 \alpha^2$. 

The most important feature of the AP model is that it allows for two 
kinds of vacuum AdS solutions. One with constant dilaton, pure AdS$_2$, 
to which we will refer as constant dilaton vacuum (CDV). The 
other one is a solution with a non-constant, linearly varying, dilaton 
which we will call linear dilaton vacuum (LDV). When uplifted to  
$(d+2)-$dimensions, these vacuum solutions produce different spacetimes. 
The CDV produces a spacetime of the form AdS$_2\times R^{d}$, i.e. an 
intrinsically 2D spacetime. On the other hand, the uplifting of the LDV 
leads to a hyperscaling violating geometry $H^{d+2}$~\cite{Dong:2012se}, 
which describes the warping of AdS$_2$ with $R^d$. In this case, 
the dilaton plays the role of the radius of $R^d$.    

The JT model has been widely used as a 2D toy model for higher-dimensional black holes, 
to give a microscopic interpretation of black entropy~\cite{Cadoni:1998sg} 
and to understand the AdS/CFT correspondence in 2D in terms of 
asymptotic symmetries~\cite{Cadoni:1999ja,Cadoni:2000gm,Cadoni:2000ah}. 
Conversely, the extended AP model has been recently used 
as a description of extremal black holes in 2D (in particular to investigate 
the breakdown of semi-classical thermodynamics and the flow from LDV 
to CDV). Moreover it has also been used to describe the back-reaction on 
holographic correlators~\cite{Almheiri:2014cka,Almheiri:2016fws} and 
to investigate its
relation with the conformal symmetry breaking~\cite{Maldacena:2016upp}.

In this paper we present a revisitation of the AP model from a 2D bulk 
perspective. The goal of our reconsinderation is twofolds. On the one hand,  
we want to describe the pattern of conformal symmetry breaking and to 
explain its dynamical consequences (generation of an IR scale and the appearance of 
Goldstone modes) focusing mainly on bulk gravitational features 
of the solutions. 
On the other hand, we would like to connect and translate the formulation 
of the boundary theory in terms of the Schwarzian action of 
Ref.~\cite{Maldacena:2016upp} in the language of 
Refs.~\cite{Cadoni:1999ja,Cadoni:2000gm,Cadoni:2000ah}, i.e. in the 
language of canonical realisation of the asymptotic symmetry group of AdS$_2$.

We will show that the pattern of conformal symmetry breaking and its 
dynamical consequences can be simply described using bulk Killing vectors, 
the covariant (bulk) mass definition of Refs.~\cite{Mann:1992yv,Gegenberg:1995jy,Cadoni:1996bn} 
and the flow between a ``symmetry-respecting'' vacuum and a ``symmetry-violating'' 
vacuum. In this way we can easily understand, from a purely 2D bulk gravitational  
perspective, the generation of an IR scale (the mass gap/scale of conformal 
symmetry breaking in the conformal correlators) and the appearance of 
local Goldstone modes. 

Our bulk perspective will also allows us to compare, energetically, the 
CDV and LDV vacuum. We will show that whereas at non-vanishing temperature 
the LDV is always energetically preferred, at $T=0$ the situation is reversed 
and the CDV is favourite. This signalises a $T=0$ quantum phase transition 
which, from an higher-dimensional perspective, can be thought as a spontaneous 
dimensional reduction from a $d+2$ to $d=2$ dimensions spacetime, 
whose possible role in quantum gravity has been emphasised in Ref.~\cite{Carlip:2012md}.

The structure of the paper is the following. In Sect.~\ref{sec:section2} 
we revisit the various solutions of the AP model. 
In Sect.~\ref{section3} we define the covariant mass 
and explain its leading role as the physical mass of the solutions. 
In Sect.~\ref{section4} we discuss the symmetries of the model, 
the pattern of the conformal symmetry breaking as well as their consequences 
for the boundary theory. In Sect.~\ref{section5} we discuss the 
free energy of the solutions and the quantum phase transition. 
In Sect.~\ref{section6} we present our conclusions. 
In the appendix~\ref{appendix} we discuss the uplifting 
of a 2D model to a $(d+2)-$dimensions model exhibiting 
hyperscaling violation in the ultraviolet (UV).

\section{ Solutions and vacua
\lb{sec:section2}}

In Schwarzschild coordinates and in absence of matter ($f=0$), 
owing to 2D Birkhoff theorem, the most general solution of the 
model~(\ref{lagrangians}) is a two-parameter family of solutions,  
\bea \label{staticmetric:eq} 
  ds^2&=& -(\lambda^2 x^2 -a^2)dt^2 + (\lambda^2 x^2 -a^2)^{-1} dx^2 ,\nonumber\\
  \eta&=& \alpha^2 + \eta_{0} \lambda x.
\eea
where $a^2$ and $\eta_0$ are dimensionless integration constants. 
The ADM mass of the solution depends on both $\eta_0$
 and $a$ and on the parameter  $\lambda$ ~\cite{Almheiri:2014cka}
\beq\lb{MADM} 
  M_{\text ADM}=\frac{\eta_{0}\lambda a^2}{2}.
\eeq 
It is important to stress that the ADM mass does not depend on 
the parameter $\a^2$ appearing in the AP potential (\ref{potential}) for the dilaton.

An important feature of the AP model is that it allows for two 
different  vacuum solutions, i.e. solutions with $M_{\text ADM}=0$. 
In fact, for $\eta_0 = 0$  and $\alpha^2  \neq 0$ we have the 
{\it constant dilaton vacuum}. It describes the $AdS_{2}$ spacetime 
with a constant dilaton. It is already well known that this vacuum does 
not allow for finite energy excitations
~\cite{Almheiri:2014cka,Almheiri:2016fws,Maldacena:1998uz}. 
It is separated from the continuous part of the spectrum by a mass 
gap, $M_{\text gap}$. This is immediately evident from Eq.~(\ref{MADM}): 
for $\eta_0=0$,  $M_{ADM}$ identically vanishes, independently from the value of $a$.
Conversely, for $\eta_0 \neq 0$ we have the {\it linear dilaton vacuum}, 
which is $AdS_2$ endowed with a linear dilaton. Differently from 
the CDV, this vacuum allows for continuous excitations with $a^2>0$. 

It is important to stress that we have two different LDV depending on 
the value of the parameters ($\eta_0, a$) and $\alpha$. For $\alpha^2=0$ 
the AP model reduces to the JT model and the solution for the dilaton is 
linear and homogeneous. Notice that this is not an exact solution of the 
AP model (i.e. the model with  $\alpha^2\neq 0$) but appears only as an 
asymptotic solution for $x\to\infty$. On the other hand, for $\eta_0 \neq 0$ 
and $\alpha^2 \neq 0$, the LDV is an exact solution of the AP model and 
interpolates between the CDV at small $x$ and a linear, homogeneous 
dilaton at large $x$. Whenever the distinction between these two LDV 
will be necessary, we will call the LDV with $\alpha^2 \neq 0$ 
{\it interpolating linear dilaton vacuum} (ILDV).   

If one uses $M_{ADM}$ as the mass of the solution the three vacuua 
become completely degenerate: the CDV simply does not allow for 
$M_{ADM}\neq 0$ excitation, whereas LDV and ILDV have the same 
energy because $M_{ADM}$ does not depend on $\alpha$.

The finite, $M_{ADM}>0$, excitations of the LDV and ILDV can 
be interpreted as 2D black holes with horizon radius 
$x=a/\lambda$ and temperature and entropy given by    
\bea 		\label{jtentropy:eq}
  &&T=\frac{\lambda a}{2 \pi},\quad 	S=2 \pi \eta_{h}= 2\pi \alpha^2 + {2 \pi \eta_0 a },\\
  &&M_{ADM}=\frac{2\pi^2\eta_0}{\lambda} T^2. \label{jtadmscaling:eq} 
\eea
 Notice that the interpretation of the 
$a^2>0$ solutions as 2D black holes is not completely 
straightforward. In fact, it is well known that the 
metric~(\ref{staticmetric:eq}) can be brought by a 
coordinate transformation in the maximally extended form 
$ds^2 = -\cosh^{2} \rho \, d\tau^2 + \frac{d\rho^2}{\lambda^2}, \,\,
-\infty<\tau,\,\rho< + \infty$, which describes full, 
geodesically complete $AdS_2$ (see e.g. Ref.~\cite{Achucarro:1993fd}). 
This is not anymore true if one takes into account 
the fact that points where the dilaton vanishes have to be 
considered spacetime singularities. This makes solutions 
with different $a^2$ as globally inequivalent and allows 
for the interpretation  of the $a^2>0$ solution as a 2D 
black hole~\cite{Cadoni:1994uf}. 
		
The previous argument forbids the existence of 2D black 
hole solutions with constant dilaton, in agreement with 
the absence of finite energy excitations of the CDV. On 
the other hand, we can formally consider zero mass thermal 
excitation of the CDV of the form 
\bea \label{staticmetric:eqCDV} 
  &&ds^2= -(\lambda^2 x^2 -\frac{4\pi^2 T^2}{\lambda^2})dt^2 + (\lambda^2 x^2 -\frac{4\pi^2 T^2}{\lambda^2})^{-1} dx^2,
  \nonumber
\\
  &&\eta= \alpha^2 .
\eea
The discussion of the spacetime singularities is much simpler 
using light-cone coordinates $x^\pm$. Using the $SL(2,R)$ isometric 
transformations, the solution (\ref{staticmetric:eq}) becomes
\bea\lb{TTT}
  &&ds^2 = -\frac{4}{\lambda^2 (x^+ - x^-)^2}dx^+ dx^-,\nonumber\\ 
  &&\eta=\alpha^2 +\frac{2\frac{\eta_0 }{\lambda}-M_{ADM} x^{+} x^{-}}{x^{+}-x^{-}}. 
\eea
The $\eta=0$ singularity is located at 
\bea
\label{singposition:eq}
  &&\left( x^{+}+\frac{\alpha^2}{M_{ADM}} \right) \left( x^{-} - \frac{\alpha^2}{M_{ADM}} \right)= \nonumber\\
  &&=\frac{\frac{2\eta_{0}}{\lambda}M_{ADM} - \alpha^{4}}{M_{ADM}^{2}},
\eea
whereas the timelike asymptotic boundary of AdS$_2$ is located 
at $x^{+}=x^{-}=t= \pm \sqrt{\frac{2\eta_{0}}{\lambda M_{ADM}}}$.
The nature of the singularity depends on the value of $M_{ADM}$. 
For $M_{ADM}> \lambda \alpha^4/(2 \eta_0)$ 
the singularity is spacelike whereas for $M_{ADM}< \lambda \alpha^4/(2 \eta_0)$ 
it is timelike.

For planar spatial topology the ILDV gives a nice, effective, 
2D description of the flow from an AdS$_2\times R^{d}$~\footnote{AdS$_2\times S^{d}$ 
in the case of spherical spacial topology} geometry in the IR to a 
hyperscaling violating geometry~\cite{Dong:2012se} in $(d+2)-$dimension 
in the UV, of which AdS$_{d+2}$ is a particular case. 
From the 2D perspective this flow is a simple consequence of both 
the relation $R\propto \eta^p$ between the dilaton and the radius 
$R$ of $R^d$ and of the constant/linear behaviour of the 
dilaton at small/large $x$. We briefly discuss the uplifting of 
the AP model to a $(d+2)$-dimensional theory with hyperscaling violation 
in the appendix~\ref{appendix}.

From the thermodynamical point of view, the CDV gives the typical 
$T=0$, extremal, state with non vanishing entropy of a large class 
of $(d+2)-$dimensional extremal black holes, like e.g. charged 
Reissner-Nordstr\"{o}m black holes in four dimensions.  
Near extremality, the mass-temperature relation for the excitations, 
$M_{ADM}\propto T^2$ in  Eq.~(\ref{jtadmscaling:eq}), implies the 
breakdown of the thermodynamical semi-classical description and 
the appearance of a mass gap \cite{Almheiri:2016fws,Preskill:1991tb}
\beq 		\label{jtmassgap:eq}
M_{gap}= \frac{\lambda}{2 \pi^2 \eta_{0}}.
\eeq
 This mass gap is the consequence of 
the absence of finite energy excitations of the CDV \cite{Maldacena:1998uz}, 
which in turns is related to the strong backreaction on AdS$_2$. 
From the AdS/CFT correspondence point of view, the appearance of the 
mass gap~\eqref{jtmassgap:eq} can be also explained in terms of 
the pattern of breakdown of the conformal symmetry which 
generates, in the IR, a mass scale of order 
$\lambda$~\cite{Almheiri:2014cka,Almheiri:2016fws,Maldacena:2016upp}.

Despite the successes of the AP model described above, 
two aspects are still not completely clear. 
The first is 
the characterisation of the energy of the solution trough 
the ADM mass~(\ref{MADM}). This mass does not distinguish 
between the different vacua of the theory, in fact it is 
independent from $\alpha$.  Due to the different 
features of the two vacua, the ADM-mass degeneracy between 
the CDV and LDV becomes particularly ambiguous. Moreover, 
$M_{\text ADM}$ does not keep informations about the presence of 
the mass gap. 
Last but not least, it also does not seem a 
suitable physical parameter to characterise the singularity. 
Indeed, the transition between spacelike and timelike 
singularities occurs not when $M_{\text ADM}$ changes sign, 
as expected, but rather at strictly positive values.

The second aspect is the characterisation of the pattern of 
conformal symmetry breaking. This pattern has been described 
using correlation functions in the dual conformal field theory. 
However, in the spirit of the AdS/CFT correspondence one should 
be able to characterise completely this pattern also using  
only the 2D gravity theory. In what follows, 
we will show how the peculiarities of dilaton gravity in 2D 
 spacetime can help us to clarify the picture presented 
above.

\section{ Covariant mass \lb{section3}}
The first peculiarity of 2D dilaton gravity is that the 
metric always admits the existence of a Killing vector whose explicit form 
depends on the dilaton~\cite{Mann:1992yv,Cadoni:1996bn,Cadoni:2000ah} 
\beq\label{kill_dil:eq}
\chi^{\mu} =F_{0} \epsilon^{\mu \nu} \partial_{\nu} \eta,
\eeq
where $F_0$ is a normalisation factor.
The second is the existence of a covariant, conserved mass
~\cite{Mann:1992yv,Gegenberg:1995jy}
\beq\label{geometricmass:eq}
M=-\frac{F_{0}}{2} \left[\int^{\eta}{V(s)ds +(\nabla \eta)^2}\right].
\eeq
In this paper we will use the normalisation prescription of 
Refs.~\cite{Cadoni:1996bn,Mann:1992yv}, i.e  $F_{0}=(\lambda \eta_0)^{-1}$.

The covariant mass $M$ gives a definition of the energy of the 
solution, which is invariant under Weyl transformation of 
the metric~\cite{Cadoni:1996bn}. It may differ from the standard 
ADM mass only by a constant (temperature independent) term.  
For this reason it is particularly appropriate to quantify 
the energy of the different vacua of the  AP model.

Computing the covariant mass for the general solution
~(\ref{staticmetric:eq}) of our model we get, 
\beq \label{jtmasscov:eq}
M= \frac{a^2 \eta_{0} \lambda}{2} - \frac{\alpha^4 \lambda}{2\eta_{0}} =
M_{ADM}  - \frac{\alpha^4 \lambda}{2\eta_{0}}.
\eeq
There are several reasons indicating that the covariant mass 
$M$ and not the ADM mass $M_{ADM}$ has to be considered as 
the physical mass of the solutions. For $\alpha= 0$ we have 
$M=M_{ADM}$. By using $M$ instead of $M_{ADM}$ we remove the 
degeneracy between the CDV and the LDV and keep also track 
about the nonexistence of finite energy excitations of the CDV.    
The ILDV has negative energy $M=- \frac{\alpha^4 \lambda}{2\eta_{0}}$, 
whereas for the CDV we have $M\to -\infty$. Moreover, the r.h.s.~of 
Eq.~(\ref{singposition:eq}) can be written as $2\eta_0 M/M_{ADM}^2$. 
Thus, the spacetime singularity is spacelike for $M>0$, whereas it becomes 
timelike for $M<0$.

\section{ Symmetries and symmetry breaking }
\lb{section4}
Let us now discuss the symmetries of the different vacua of the 
AP model. The isometry group of AdS$_2$ is the $SL(2,R)\sim SO(1,2)$ 
group generated by three Killing vectors. In Schwarzschild 
coordinates~(\ref{staticmetric:eq}) they represent time translations 
${\cal T}$, dilatations and special conformal transformations. 
However, the $SL(2,R)$ symmetry is only a symmetry of 
the metric. The whole solution contains also the dilaton
which, under isometric transformations generated by the Killing
vector $\chi$, transforms as 
$\delta \eta= \L_{\chi} \eta= \chi^{\mu} \partial_{\mu} \eta$ ~\cite{Cadoni:2000ah}.
Notice that a constant dilaton will preserve the $SL(2,R)$ symmetries 
of the metric, whereas a non constant dilaton will necessarily 
break explicitly the $SL(2,R)$ symmetry. 
On the other hand, the 2D metric  allows 
for the killing vector~(\ref{kill_dil:eq}), which is also always 
a symmetry of the dilaton ($\delta \eta=0$). 
Thus, a non constant 
dilaton breaks explicitly the full $SL(2,R)$ symmetry group of 
AdS$_2$ down to its subgroup $H$ generated by the Killing 
vector~(\ref{kill_dil:eq}).

In the case of the static solutions~(\ref{staticmetric:eq}) the 
residual symmetry is the time translations ${\cal T}$ and the symmetry 
breaking pattern is $SL(2,R)\rightarrow {\cal T}$~\cite{Cadoni:2000ah}.
As a consequence, the CDV of the AP model preserves the full $SL(2,R)$ 
symmetries of AdS$_2$, whereas the LDV breaks $SL(2,R)\rightarrow {\cal T}$. 
In this way we can describe the IR/UV flow CDV$\rightarrow$LDV as a 
symmetry breaking of the full $SL(2,R)$  group down to time translations.

The parameter controlling the symmetry breaking is $\partial_{x} \eta=\eta_{0} \lambda$.  
Any $\eta_0\neq 0$ value breaks the $SL(2,R)$ symmetry to ${\cal T}$ 
and generates a mass-scale in the IR, set by $\lambda$, which is 
of the same order of magnitude of the mass gap~(\ref{jtmassgap:eq}).

Thus, the presence of a non constant dilaton breaks 
the conformal symmetry of the AdS$_2$ background and generates 
a mass gap through $\eta_0\neq0$. Further, it also affects the 
asymptotic symmetries of AdS$_2$~\cite{Cadoni:2000ah} 
and the dynamics of the boundary theory. In particular, 
the latter can be  constructed using boundary curves $t(u)$, 
where $u$ is the time coordinate in the one-dimensional 
regularised boundary of AdS$_2$~\cite{Maldacena:2016upp}. 

Actually, the two descriptions, that of Refs.~\cite{Cadoni:1999ja,Cadoni:2000ah}, 
which uses canonical realisation of the asymptotic symmetry 
group (ASG) of AdS$_2$ and that of Ref.~\cite{Maldacena:2016upp} 
give similar results but using different languages and a 
different coordinate system.

The ASG of AdS$_2$~\cite{Cadoni:2000gm,Cadoni:2000ah} is given 
by reparametrisations of the type $\xi^{t}=\epsilon(t), \quad \xi^{x}=x\epsilon'(t)$ 
and is generated by a single copy of the Virasoro algebra. 
This transformations map the boundary curves 
$t(u)$ into curves  $t(u)=u+\epsilon(u)$.
On the other hand, they act on the asymptotic expansion of the 
metric  \cite{Cadoni:1998sg,Cadoni:1999ja,Cadoni:2000ah}
\bea\lb{lkj}
&&g_{tt} = -\lambda^2 x^2 + \gamma_{tt} (t)+ o(x^{-2}),\\
&& g_{tx}= \frac{\gamma_{tx}(t)}{\lambda^3 x^3} + o(x^{-5}),\\
&&g_{xx} = \frac{1}{\lambda^2 x^2}+ \frac{\gamma_{xx}(t)}{\lambda^4 x^4} +
o(x^{-6}),
\eea
by transforming the values of the boundary fields $\gamma$.

In the case of the CDV (which is called ``pure AdS$_2$'' in Ref.
\cite{Maldacena:2016upp}), the dilaton is constant, $\eta_0=0$, 
and we do not have the explicit breaking of the $SL(2,R)$ conformal 
symmetry. The full Virasoro ASG is 
spontaneously broken by the AdS$_2$ bulk geometry down to 
the $SL(2,R)$ group of isometries. The zero modes can be 
characterised either by the boundary curves $t(u)$ (in the 
language of Ref. \cite{Maldacena:2016upp}) or by the boundary 
deformations $\gamma$ (in the language of Ref.~\cite{Cadoni:2000gm,Cadoni:2000ah}). 
These zero modes can be viewed as the Goldstone modes associated 
to the spontaneous breaking of the ASG~\cite{Maldacena:2016upp}. 
However, these modes are not local, there is no local action one 
can write for them and this is related to the fact that the 
central charge $c$ in the Virasoro algebra is zero.

In the case of the  LDV (which is called "nearly AdS$_2$" in 
Ref.~\cite{Maldacena:2016upp}), as discussed above, the non 
constant value of the dilaton breaks explicitly $SL(2R)\to \cal{T}$. 
This gives, in the language of Ref.~\cite{Maldacena:2016upp} a 
new dimensional coupling constant, the renormalised boundary 
value of the dilaton $\phi_r(u)$, which can be used to 
constrain the shape of $t(u)$ and to produce a local, 
Schwarzian action for the pseudo-Goldstone bosons $t(u)$.	
	 
Conversely, in the language of Refs.~\cite{Cadoni:2000gm,Cadoni:2000ah}, 
the explicit breaking of the conformal symmetry is described by 
the asymptotic expansion for the  dilaton $\eta=\eta_{0}  \rho(t) x + 
+ o(x^2)$ with the boundary field $\rho(t)$ transforming as 
$\delta{\rho}= \epsilon \dot{\rho} + \dot{\epsilon} \rho$ 
under the action of the ASG~\cite{Cadoni:2000gm,Cadoni:2000ah}.

The two boundary fields can be identified: $\phi_r(u)=\eta_0\rho(t(u))$. 
In both descriptions the physical effect of the explicit 
symmetry breaking is to make the Goldstone modes local and 
to generate a non vanishing central charge in the Virasoro 
algebra,
\beq\lb{cc}  
c=12 \eta_0,
\eeq
through the anomalous transformation of the boundary stress 
energy tensor $T_{tt}$ under the action of the ASG. In fact, we have 
$T^{(1)}_{tt}= \phi_r\{t(u),u\}$ for the boundary theory of 
Ref.~\cite{Maldacena:2016upp}, whereas $T^{(2)}_{tt}= -2 \eta_0/\lambda \ddot \rho$ 
for the boundary theory of Refs.~\cite{Cadoni:2000gm,Cadoni:2000ah}. 
The central charge $c$ takes the form given by Eq.~\eqref{cc} 
if we choose $\rho=1$. We can always make this choice by fixing 
the $u$-reparametrisation in the boundary which corresponds, 
in the language of Refs.~\cite{Cadoni:2000gm,Cadoni:2000ah},
to consider deformations of the dilaton near the on-shell 
solution (see~\cite{Cadoni:2000gm,Cadoni:2000ah} for details).

The stress energy tensor $T^{(2)}_{tt}$ can be brought in the 
form $T^{(1)}_{tt}$. In fact, by considering finite transformations 
associated with the infinitesimal ones characterised by $\epsilon=u$, by 
using the transformation of the boundary field $\rho$ and 
by setting $\rho=1$ one finds $T^{(2)}_{tt}= (c/12)\{t(u),u\}$.  The
link between the origin of Schwarzian action and the 
presence of a non constant dilaton
was emphasised also in Ref. \cite{Cvetic:2016eiv}, where it was shown 
 that in a holographic framework the effective action
of the AP model can be put in a Schwarzian form
using the anomalous trace Ward identity. In particular,
the anomaly turns out to be proportional to the
source of the scalar operator dual to the dilaton, which  
is  the analogue of our function $\rho(t)$. 

Summarising, the explicit breaking of the conformal symmetry, 
$SL(2,R)\rightarrow {\cal T}$ generated by a non-constant 
dilaton has two effects. First, it generates an IR scale in 
the form of mass gap, $M_{\text gap}\sim \lambda$, separating 
the CDV from the LDV.

 Second, it transforms the global 
Goldstone modes of the CDV associated with the spontaneous 
breaking of the ASG into local pseudo-Goldstone modes, 
producing a central charge $c=12\eta_{0}$ in the Virasoro 
algebra associated to the ASG. This central charge therefore 
counts the number of pseudo-Goldstone modes. From this 
perspective we can identify the degrees of freedom 
responsible for the entropy of the 2D dilatonic black 
hole as these pseudo-Goldstone modes. The microscopic 
derivation of the entropy of the 2D dilatonic black hole 
given in Ref.~\cite{Cadoni:1999ja} can be seen as counting 
the states of these modes. It is also interesting to notice 
that the mass gap in the chiral 2D CFT can be also understood 
as finite size effect generated by a plane/cylinder 
transformation of the vacuum of a CFT with non vanishing 
central charge $c=12 \eta_0$~\cite{Cadoni:2000fq,Cadoni:2001tb}.

\section{ Quantum phase transition and spontaneous dimensional reduction}
\lb{section5}

The AP model allows for two different class of solutions, 
namely the 2D black hole~(\ref{staticmetric:eq}) and the 
zero mass thermal excitations of the CDV~(\ref{staticmetric:eqCDV}). 
One important question is to determine which of these 
two solutions is, from the thermodynamic point of view, 
globally favourite. 
%
 Using the 2D Hamiltonian formalism~\cite{Caldarelli:2000xk}, 
this can be done by computing the difference  $\Delta F$ between the 
free energy, $F$, of the two solutions. 
In the case under consideration this computation 
is not straightforward because $\Delta F$ is usually 
computed for solutions having the same asymptotical behaviour.

The presence of the dilaton makes the asymptotics 
of the two classes of solutions of the AP model (linear and 
constant dilaton, respectively) different, thus preventing 
the standard computation of $\Delta F$.  
This problem can be circumvented  by defining the free energy 
of the solution  with respect to its own 
vacuum~\cite{Cadoni:2012uf}. This method has been applied, 
for example, in Ref.~\cite{Cadoni:2012uf} to calculate 
$\Delta F$ for two classes of 4D solutions approaching 
asymptotically to AdS and to a solution with hyperscaling 
violation, respectively.

Using this prescription for the free energy $F$ in the 
Euclidean action formalism, for the case under consideration 
we get:
\bea
\label{LDFreeEnergy}
  F^{BH}&=&-\frac{2\pi^2\eta_0}{\lambda}T^2-2\pi\alpha^2 T,\nonumber\\
  F^{T}&=&-\kappa\eta_{h}^C=-2\pi\alpha^2T,\\
  \Delta F &=& F^{BH}-F^{T}= -\frac{2\pi^2\eta_0}{\lambda}T^2\nonumber,
\eea
where $F^{BH}$ is the free energy of the 2D black hole~
(\ref{staticmetric:eq}) obtained by subtracting the 
contribution of the ILDV, whereas $F^{T}$ is the free 
energy of the thermal excitation of the CDV obtained by 
subtracting the contribution of its own vacuum. From 
Eq.~(\ref{LDFreeEnergy}) follows immediately that for 
any $T\neq 0$, $\Delta F<0$ and the 2D dilatonic black 
hole is energetically preferred.

By construction, Eq.~(\ref{LDFreeEnergy}) does not give 
any information about the behaviour at $T=0$, being $F$ 
defined with reference to the respective vacua at $T=0$.  
Formally, at $T=0$ the two vacua are degenerate, consistently 
with the degeneracy of the CDV and ILDV when the ADM mass 
is used to characterise the two solutions. 

Moreover, the semi-classical approximation, on which the 
euclidean action formalism is based, breaks down at 
$T\sim M_{\text gap}$, so that Eq.~(\ref{LDFreeEnergy}) 
cannot be trusted at $T=0$.

At $T=0$ there is no thermal contribution to the free 
energy and $\Delta F$  is given by the mass difference 
between the two vacua, $\Delta F_{T=0}=\Delta M=M^{ILDV}-M^{CDV}$.
 
We have already argued that we should use the covariant 
mass~(\ref{jtmasscov:eq}) as the physical mass instead 
of the ADM mass.
Using this mass in the computation we find $\Delta F_{T=0}\to \infty >0$. 
This means that the CDV is energetically preferred and 
that at $T=0$ the 2D dilatonic black hole undergoes a 
quantum phase transition to the CDV. The free energy of 
the CDV diverges. 
Obviously, this is a manifestation of the presence of a 
mass gap in a classical theory. We expect a full quantum 
treatment of the topic to remove this divergence but not 
to change the  $\Delta F_{T=0}$ result.

From a four-dimensional perspective this quantum phase 
transition can be interpreted as a spontaneous dimensional 
reduction. In fact, the $(d+2)-$dimensional uplifting of 
the ILDV is a scale covariant geometry $H^{d+2}$ with 
hyperscaling violation in $(d+2)-$dimensions, whereas the 
uplifting of the CDV is AdS$_{2}\times R^{d}$, i.e. a geometry 
which is intrinsically two-dimensional, being the radius of 
$R^{d}$ not dynamical. In terms of the uplifted geometries 
we have the $T=0$  phase transition 
$H^{d+2}\to$ AdS$_{2}\times R^{d}$.

This  phase transition supports the suggestion of 
Ref.~\cite{Carlip:2012md} about the existence of a spontaneous 
dimensional reduction of the spacetime to two dimensions 
near the Planck scale. 

Let us conclude with some remarks about one loop corrections 
to the free energy~\eqref{LDFreeEnergy}. Our calculation is based 
on the semi-classical approximation. One loop corrections to $F$ 
have been shown in Ref.~\cite{Maldacena:2016upp,Almheiri:2014cka}
 to have the typical 
$\log T$ behaviour, which gives a dangerous divergent term in 
the IR. However, this term does not contribute to the entropy of 
the CDV~\cite{Maldacena:2016upp}, we therefore expect our result 
to extend also beyond the semi-classical approximation.

\section{Conclusions}
In this paper we have revisited the AP dilaton gravity model 
focusing mainly on bulk features of the model. Using a 
covariant definition of the mass, bulk Killing vectors and 
the flow between the two different vacua of the theory 
characterised, respectively, by a constant and linear varying 
dilaton, we have given a description of the pattern of 
conformal symmetry breaking, which is complementary to that 
emerging in the dual CFT~\cite{Maldacena:2016upp}. 

This pattern is quite similar to that pertinent to hyperscaling-violating geometries in higher dimensions, 
to which we show the AP model can be uplifted. In fact, as a 
result of the flow between a ``symmetry-violating'' vacuum and 
a ``symmetry-respecting'' vacuum an IR scale is generated in 
the form of a mass gap. The other effect of the conformal 
symmetry breaking is to make local the Goldstone modes 
associated with the asymptotic symmetries of the 2D spacetime. 
This generates a non-vanishing central charge in the dual conformal 
theory, which explains at microscopic level the entropy of the 
2D black hole~\cite{Cadoni:1999ja}.

We have also shown that several features of the boundary theory 
described in Ref.~\cite{Maldacena:2016upp} can be easily translated 
in our language, which is based on bulk quantities and on the 
asymptotic symmetries of the spacetime. 

Finally, the use of the covariant mass as measure of the energy 
of the solutions, allowed us to compare energetically the two 
different vacua of the theory, showing the existence of a zero 
temperature phase transition in which the vacuum with constant 
dilaton is energetically preferred. We speculate that this 
quantum phase transition could be related to the spontaneous 
dimensional reduction of the spacetime to two dimensions near 
the Planck scale described in Ref.~\cite{Carlip:2012md}.

\lb{section6}
\section {Appendix}
\label{appendix}
Let us briefly show how the solution~(\ref{staticmetric:eq}) 
can be uplifted to a $(d+2)-$dimensional geometry describing the 
flow from a AdS$_2\times R^d$ geometry in the IR to a 
$(d+2)-$dimensional geometry with hyperscaling violation in 
the UV.

In $(d+2)-$dimensions the model is described by the action
\begin{equation}
\lb{dr}
S= \int d^{d+2} x \sqrt{-g_{(d+2)}} R_{(d+2)} + {\cal L}_{M},
\end{equation}
where $L_M$ is the Lagrangian for matter fields, which may 
also contains explicit coupling of matter fields to the dilaton.
 
For simplicity we assume that after dimensional reduction to 
two spacetime dimensions, the term ${\cal L}_{M}$ either 
reproduce exactly the potential~(\ref{potential}) or a potential, 
which can be approximated by~(\ref{potential}). We look for brane 
solutions of the model, i.e. solutions for which the $d$-dimensional 
spatial sections have planar topology $R^d$ 
 and the dilaton, 
$\eta$, plays  the role of the radius: 
$ds^2_{(d+2)}=ds^2_{(2)}+ \eta^{2/d}dx_idx^i$. 
In general, the dimensional reduction of the action~(\ref{dr}) on 
this background produces kinetic terms for the dilaton $\eta$ in the 
2D dilaton gravity action. This terms can be put to zero by a Weyl 
rescaling of the 2D metric~\cite{Cadoni:1996bn}. This corresponds to 
use, instead, the dimensional reduction
\beq\lb{dr1}
  ds^2_{(d+2)}=\eta^{\frac{1-d}{d}}ds^2_{(2)}+ \eta^{\frac{2}{d}}dx_idx^i.
\eeq
One can check that the dimensional reduction now produces the 
AP action~(\ref{2dlagrangian:eq}). Using  Eq.~(\ref{dr1}) and the 
form of the 2D solution given by Eq.~(\ref{staticmetric:eq}), 
one can easily realise that the $(d+2)-$dimensional solution 
interpolates between an hyperscaling violating geometry at 
large $x$ and an AdS$_2\times R^{d}$ geometry at small $x$. 
In fact, for $x\to \infty$ the term proportional to $x$ in the dilaton 
dominates, and the change of radial coordinate, $x\propto r^{-2d/(d+1)}$, 
brings the metric in the scale covariant form given in Ref.~\cite{Dong:2012se}   
\beq\lb{dr2}
  ds^2_{(d+2)}=r^{-2\frac{d-\theta}{d}}\left(-r^{-2(z-1)}dt^2+dr^2+dx_idx^i\right).
\eeq
The hyperscaling violating parameter $\theta$ and the 
dynamical exponent $z$ are:
\beq
\theta= \frac{d (d-1)}{d+1},\quad z= \frac{2d}{d+1}.
\eeq
Conversely, in the near horizon limit the term proportional to $x$ in the 
dilaton, can be neglected with respect to the constant term and 
the metric~(\ref{dr1}) gives an AdS$_2\times R^{d}$ geometry. 
This can be also considered as the limiting case $\theta=0,\, z=\infty$ 
of the hyperscaling violating geometry~(\ref{dr2}).


%

\end{document}